    \documentstyle[11pt]{article}
    
    \newcommand{\be}{\begin{eqnarray}}
    \newcommand{\ee}{\end{eqnarray}}
    \newcommand{\eel}[1]{\label{#1}\end{eqnarray}}
    \newcommand{\bes}{\begin{eqnarray*}}
    \newcommand{\ees}{\end{eqnarray*}}
    
    \newcommand{\hb}{{\cal i}}

    \newcommand{\ga}{{\gamma}}
    \newcommand{\la}{{\lambda}}

    \newcommand{\pa}{\partial}
    \newcommand{\ra}{{\rightarrow}}
    
    \newcommand{\cG}{{\cal G}}
    \newcommand{\cD}{{\cal D}}
    
    \newcommand{\cF}{{\cal F}}
    \newcommand{\cH}{{\cal H}}
    
    \newcommand{\cL}{{\cal L}}
    \newcommand{\cP}{{\cal P}}

    \newcommand{\halv}{\frac{1}{2}}
    \newcommand{\nn}{\nonumber}
    
\begin{document}
    \def\uslash{{U\mskip -12mu/}}

    \begin{quote}
    \center{\Large{\bf Role of Zero Modes in
    Quantization of QCD in Light-Cone Coordinates}}
    \vskip4ex
    \center{{\sf Hiroyuki Fujita} and {\sf Sh. M. Shvartsman} }
    \vskip2ex
    \center{{\sl Department of Physics, Case Western Reserve University, \\
    Cleveland, Ohio 44106}}
    \vskip5ex
    \end{quote}
    \bigskip\bigskip

    \begin{abstract}

    	Two-dimensional heavy-quark QCD is studied in
    the light-cone coordinates with (anti-) periodic field boundary conditions.
    We carry out the light-cone  quantization of
    this gauge invariant model.  To examine the role of the zero modes of
    the gauge degrees of freedom,
    we consider the quantization procedure in the zero mode
    and the nonzero mode sectors separately. In both sectors, we obtain the
    physical variables and their canonical (anti-) commutation relations.
    The physical Hamiltonian is constructed via a step-by-step elimination
    of the unphysical degrees of freedom. It is shown that the
    zero modes play a crucial role in the self-interaction
    potential of both the heavy-quarks and gluons, and in the interaction
    potential between them. It is also shown that
    the Faddeev-Popov determinant depends on the zero modes of the gauge
degrees of
    freedom. Therefore, one needs to introduce the Faddeev-Popov ghosts
    in their own nonzero mode sector.
    \end{abstract}
    \vspace{1.75in}

    \newpage
    \section{Introduction}

    \hspace{3em}One of the promising approaches to problems in
    QCD is the light-cone quantization \cite{brev1,brev2}.
    The light-cone quantization has turned out to be a useful tool
    for the perturbative treatment of field theories \cite{Brodsky,Brodsky1}.
    In its extension [9-13] to the
    nonperturbative domain, one has come to realize that careful
    attention must be paid to
    the nontrivial vacuum structure of the light-cone quantum field theory.
    Some mathematical aspects of the question with regard to the existence of
    such vacuum states were considered
    in \cite{Schlieder,Coester}. For instance, the
    light-cone vacuum in the massless Schwinger model can be only understood
    by  careful study of the zero modes
    of the constraints imposed by the light-cone frame
    \cite{mccartor,Heinzl}.  Indeed, it has
    been conjectured that the dynamics of the zero modes in QCD in
    the light-cone quantization provides the mechanism for the confinement
    \cite{brev1,brev2}.

    	In the present paper, we continue the quantization of
    heavy-fermion gauge theory started in our previous article \cite{Brown}.
    We apply the Faddeev-Jackiw quantization technique to the two-dimensional
    heavy-fermion QED and two-dimensional heavy-quark QCD. Although in this
case
    the quantization of the full model (not only in heavy mass limit but
    also in light mass limit) can be
    performed, we restrict ourselves to heavy-fermions.
    We carry out the quantization of these
    models in a light-cone domain restricted in its
    ``spatial'' directions. It is well known that in such a restricted
    region one has problems with the zero modes
    \cite{Maskawa,Heinzl} which, as it was mentioned above,
     turned out to be the most important
    variables in this case.
    The quantization of QED and QCD on the
    finite dimension manifolds (circle, torus) were considered in
    \cite{Luscher,Rajeev,Hetrick}. The role of the zero modes in QED using the
Lagrange approach was studied in \cite{kal1}.
    The dynamics of zero modes in the two-dimensional QCD employing the
light-cone
    variables was considered in \cite{kal2,Thomas,Dhar,Tachibana}.

    	In studying the quantization of the gauge field theories,
    one is confronted by first-class constraints, and, for  this reason, the
    corresponding gauge conditions should be imposed.
    A consistent canonical quantization formalism for  such problems was
    proposed by Dirac \cite{Dirac} and Bergmann \cite{Bergmann}, and its
    generalization to fermionic (Grassmann-odd) constraints by Casalbuoni
    \cite{Casalbuoni}. There is another approach to the quantization of the
    gauge theories  proposed by Faddeev and Jackiw \cite{FJ}.

    	The specific gauge theories (QED and QCD) we address
    in this paper are in terms
    of the light-cone variables where, as we will see, the quantization faces
    with some constraints involving the zero mode variables which
    require special attention.  To examine explicitly the role of the zero
modes,
    we consider the
    quantization procedure in the separated nonzero mode and zero mode
    sectors. In such sectors, we choose special gauge conditions
    in order to gauge out the nonzero modes and obtain
    the physical variables and their canonical (anti-) commutation relations.
The
    physical Hamiltonian is constructed by systematic elimination of the
    nondynamical variables.

    	The paper is organized as follows. In Section 2, the
    Faddeev-Jackiw technique in terms of
    the light-cone coordinates is applied to the two-dimensional heavy-fermion
QED.
    Despite that the results we obtain here are trivial, this example is useful
    for better understanding what is going on in QCD.  The
    Faddeev-Jackiw quantization algorithm is equivalent to
    the Dirac one, of course, but it is sometimes simpler to be employed,
    especially when the constraints are
    linear.  In  Section 3, we consider the light-cone quantization of
    the two-dimensional heavy-quark QCD, where the
    zero modes and nonzero modes of the  gauge degrees of freedom are taken
into
    account. It is shown that the physical degrees of freedom are the
    zero modes of gauge fields ($A^{(P)}_{-}$ in QED and $A^{a (P)}_{-}$
    in QCD), their conjugate momenta and the fermionic
    variables. We found that the careful elimination of
    unphysical gauge degrees of freedom leads to additional terms in
    the physical Hamiltonian. Such a Hamiltonian is constructed, and, in
    particular, the interaction potential between heavy-quarks, as well as
    the interaction Hamiltonian between heavy-quarks and gluons
    are found. It is shown that in the case of
    QCD one needs to introduce the nonzero modes of the Faddeev-Popov ghosts.

    \vskip2ex

    \setcounter{equation}{0}
    \section{2D Heavy Fermion QED}
    \vskip2ex

    \hspace{3em}In this Section, we are going to consider a
    simple example which illustrates
    the Faddeev-Jackiw quantization technique \cite{FJ} for the case
    of the 2D heavy-fermion QED when one needs to take
    the zero modes of the gauge degrees of freedom into consideration.
    This will be the  first step
    toward the attack of the QCD model, so we can kill two birds by one stone.

    	Following the Isgur and Wise \cite{Isgur}, the
    Lagrange density of the $2D$ heavy-fermion QED has the following form
    \be
    \cL=i\overline{\Psi}\uslash {U}^{\mu}D_{\mu}\Psi-
    M\overline{\Psi}\Psi+
    \halv F_{03}^2
    \eel{2.1}
    where the Minkowski metric is: ${\rm diag}\;\; g_{\mu \nu}=(1,-1)$,
    \be
    D_{\mu}=\pa_{\mu}+ieA_{\mu}
    \eel{2.2}
    is the covariant derivative,
    \be
    \uslash =\gamma\cdot U =\ga^\mu U_\mu
    \eel{2.3}
    and $U^{\mu}$ is a given
    2-velocity of the heavy-fermions subject to the condition $U^{2}=1$.
    The field strength tensor is
    \be
    F_{03}=\pa_{0}A_{3}-\pa_{3}A_{0}
    \eel{2.4}
    and we use the system of units where $\hbar=c=1$.
    The heavy-fermion limit means that the quantity $MU^{\mu}$ is
    greater than any other momenta in the problem under consideration.

    	The light-cone coordinates in the two-dimensional space  are
    $x^{\mu}=(x^{+},x^{-})$,  where
    \be
    x^{\pm}=\frac{1}{\sqrt{2}}(x^{0}\pm x^{3})
    \eel{2.5}
    The variable $x^{+}$ plays the role of the ``time'' variable, and $x^-$
    is the spatial one.
    In terms of such coordinates, the Lagrange density $\cL$  becomes
    \be
    \cL&=&i\overline{\Psi}\uslash
    ({U}_{+}\pa_{-}+{U}_{-}\pa_{+})\Psi-M\overline{\Psi}\Psi
    +\frac{1}{2}F^{2}_{+-}\nn \\
    &&-e\overline{\Psi}\uslash ({U}_{+}A_{-}+
    {U}_{-}A_{+})\Psi
    \eel{2.6}
    where
    \be
    F_{+-}&=&\pa_{+}A_{-}-\pa_{-}A_{+}\;\;,
    \;\;A_\pm =\frac{1}{\sqrt{2}}(A_0 \pm A_3)\;,\nonumber \\
    \pa _\pm &=&\frac{\pa }{\pa x^\pm}\;\;,\;\;
    U_\pm = \frac{1}{\sqrt{2}}(U_0 \pm U_3)
    \eel{2.6a}

    	The Lagrange density (\ref{2.1}) is gauge invariant.
    This means that the classical theory contains ``first-class'' constraints,
    and one needs a quantization prescription for  systems with constraints
    such as that provided, for example, by Dirac \cite{Dirac}, Bergmann
    \cite{Bergmann}, Casalbuoni \cite{Casalbuoni} or Faddeev-Jackiw
    \cite{FJ}.

    	The infrared problems  can be
    regularized by considering the system to be contained in a finite
    volume.  We consider, for this reason, the quantization of
    the theory in the
    restricted region $-L\leq  x^{-}\leq  L$,
    and impose periodic boundary conditions for
    the bosonic variables and antiperiodic boundary conditions for the
    fermionic ones
    \be
    &&A_\mu (x^+ ,x^{-}- L)=A_\mu (x^+ ,x^{-}+L),\nn\\
    &&\Psi(x^+ ,x^{-}-L )=-\Psi(x^+ ,x^{-}+L),\;\;\bar{\Psi}(x^+ ,x^{-}-L)
    =-\bar{\Psi}(x^+ ,x^{-}+L)
    \eel{2.6b}
    We choose the antiperiodic boundary conditions for the fermions in order
    to avoid their own zero mode problem,
    and treat them as Grassmann-odd variables.
    The periodic boundary conditions for the gauge variables are useful because
    when integrating by parts the boundary term vanishes. Although we choose
    antiperiodic boundary conditions for the fermions, this does not have to
    confuse anyone because the fermions  always appear in a bilinear
combination
    which is a periodic function.
    Each boundary condition implies that there is an additional
    constraint to be satisfied.  In this paper, the periodic constraints are
    imposed  explicitly. In \cite{joon} the boundary
    conditions were treated as additional constraints.

    	To start with quantization procedure, we  define
    the  momenta $\Pi_{\Psi}$ and $ \Pi_{\overline{\Psi}}$  conjugate to
    the fermionic variables $\Psi$ and $\overline{\Psi}$, and the momenta
    $\Pi_{\pm}$ conjugate to the gauge degrees of freedom $A_{\pm}$
    \be
    & &\Pi_{\Psi}=\frac{\pa_{r}\cL}{\pa \dot{\Psi}}=
    i\overline{\Psi}\uslash {U}_{-},\;\;
    \Pi_{\overline{\Psi}}=\frac{\pa_{r}\cL}
    {\pa \dot{\overline{\Psi}}}=0\;,\nn\\
    && \Pi_{\pm}=\frac{\pa \cL}{\pa \dot{A_{\pm}}}
    \eel{2.7}
    where the label ``$r$" denotes the right derivative, and the ``dot" always
    means the derivative with respect to the ``time" $x^{+}$.

    	The zero modes and nonzero modes  for the
     gauge degrees of freedom are defined  as
    \be
    A^{(P)}_{\pm}=P\ast A_{\pm}\;\;,\;\;A^{(Q)}_{\pm}=Q\ast A_{\pm}
    \eel{2.8a}
    where $P$ and $Q$ are the projection operators onto the zero mode sector
    ($P$ sector) and nonzero mode sector ($Q$ sector), respectively
\cite{Heinzl}
    \be
    P(x,y)&=&\frac{1}{2L}\;\;,\;\;Q(x,y)=\delta(x^- -y^-)-P(x,y)\;, \nn \\
    (P\ast f)(x)&=&\frac{1}{2L}\int_{-L}^{L}f(x)dx
    \eel{2.9}
    The momenta
    \begin{eqnarray*}
    && \Pi_{\pm}^{(P,Q)}\equiv \left(\Pi^{(P,Q)}_{A^{(P,Q)}_{\pm}}\right)
    \end{eqnarray*}
    conjugate to the variables $A_{\pm}^{(P,Q)}$ are
    \be
    && \Pi_{\pm}^{(P)} =P\ast
    \frac{\pa \cL}{\pa \dot{A}_{\pm}}\;\;,\;\;
    \Pi_{\pm}^{(Q)} =Q\ast
    \frac{\pa \cL}{\pa \dot{A}_{\pm}}
    \eel{2.8}
    One then obtains
    \be
    && \Pi^{(P,Q)}_{+}=0\;\;,\;\;
    \Pi^{(P)}_{-}=\dot{A}^{(P)}_{-}\;\;,\;\;
    \Pi^{(Q)}_{-}=\dot{A}^{(Q)}_{-}-\pa_{-}A^{(Q)}_{+}
    \eel{2.10}
    The velocities $\dot{A}^{(P,Q)}_{-}$ can be expressed through the momenta
    $\Pi^{(P,Q)}_{-}$, whereas $\dot{A}^{(P,Q)}_{+}$ can not. Consequently,
    one has four
    primary constraints ${\bf \Phi}= 0$
    \be
    {\bf \Phi}=\left\{ \begin{array}{ll}
    		 \phi^{(P,Q)}_{+}=\Pi^{(P,Q)}_{+}\\
    		 \phi_{\Psi}=\Pi_{\Psi}-i\overline{\Psi}
    		 \uslash {U}_{-}\\
    		 \phi_{\overline{\Psi}}=\Pi_{\overline{\Psi}}
    		 \end{array}
    	 \right.
    \eel{2.11}

    	In terms of the variables $A^{(P,Q)}_{-},\;\Pi^{(P,Q)}_{-}
    ,\;\Pi_{\Psi}$, and $\Psi$,
    the Lagrange density $\cL $ on the constraints surface
    (\ref{2.11}) can be rewritten in the Hamilton form
    \be
    \cL&=&\Pi^{(P)}_{-}\dot{A}^{(P)} _{-}+\Pi ^{(Q)}_{-}
    \dot{A}^{(Q)} _{-}+\Pi_{\Psi}\dot{\Psi} -\cH, \nn \\
    \cH&=& \cH_{F}+\cH_{EM},\nn \\
    \cH_{F}&=&-\frac{U_{+}}{U_-}\Pi_{\Psi}\pa_{-}\Psi-i\frac{M}{U_{-}}
    \Pi_{\Psi}\uslash \Psi\;, \nn \\
    \cH_{EM}&=&-ie\Pi_{\Psi}{U}^{-1}_{-}
    \Psi\left\{U_{-}A^{(P)}_{+}+U_{-}A^{(Q)}_{+}
    +U_{+}A^{(P)}_{-}+U_{+}A^{(Q)}_{-}\right\}\nn\\
    &&+\frac{1}{2}\left(\Pi^{(P)}_{-}\right)^2+\frac{1}{2}
    \left(\Pi^{(Q)}_{-}\right)^{2}
    +\Pi^{(Q)}_{-}\pa_{-}A^{(Q)}_{+}
    \eel{2.12}
    We neglect the term
    $\Pi^{(P)}_{-} \Pi^{ (Q)}_{-}$ in the Hamilton density $\cH_{EM}$
    because it does not give a
    contribution to the corresponding Hamiltonian.

    	We now start the quantization procedure following \cite{FJ}. (We do not
present here the details of this algorithm. The readers can find
    them in the original paper \cite{FJ}
    ).
    To start with, we should define the initial set of variables
    $\zeta^{(0)}_j$ and the corresponding ``generalized'' momenta
    $a^{(0)}_j(\zeta^{(0)})$ which, in the case under consideration,
    are found to be
    \be
    &&\zeta^{(0)}_{j} =
\left\{A_{-}^{(P)}\;,\;\Pi^{(P)}_{-}\;,\;A^{(P)}_{+}\;,\;
    A_{-}^{(Q)}\;,\;\Pi^{(Q)}_{-}\;,\;A^{(Q)}_{+}\;,\;\Psi\;,\;\Pi_{\Psi}\right
    \}, \nn \\
    && a^{(0)}_{j}=\left\{\zeta^{(0)}_2\;,\;0\;,\;0\;,\;\zeta^{(0)}_5\;,\;
    0\;,\;0\;,\;\zeta^{(0)}_8\;
    ,\;0\;\right\}\;\;,\;\;j=1,\ldots,8
    \eel{2.13}
    One of the most important objects in the Faddeev-Jackiw approach is the
    simplectic supermatrix $f^{(0)}_{ij}$, which is defined in general by
    \cite{Govaerts}
    \be
    f^{(0)}_{ij}=\frac{\pa_{\ell}a^{(0)}_{j}}{\pa\zeta^{(0)}_{i}}-
    (-1)^{\epsilon_{a}\epsilon_{\zeta}}
    \frac{\pa_{\ell}a^{(0)}_{i}}{\pa\zeta^{(0)}_{j}}
    \eel{2.14}
    where $\epsilon_{a}(\epsilon_\zeta)$ is the Grassmann parity of $a
    (\zeta)$
    \be
    \epsilon_{\zeta}=\left\{ \begin{array}{c}
    0\;\;,\;\;{\rm if}\;\; \zeta\;\; {\rm  is \;Grassmann-even}\\
    1\;\;,\;\;{\rm if}\;\; \zeta\;\; {\rm is\; Grassmann-odd}
    \end{array}
    \right.
    \eel{2.15}

    The simplectic supermatrix corresponding to the set of the variables
    (\ref{2.13}) is block-diagonal
    \be
    {\rm block\; diag}\;f^{(0)}_{ij}(x^-,y^-)=({\cal A},0,{\cal A}
    \delta(x^- -y^-),0,
    {\cal A}_1 \delta(x^- -y^-))
    \eel{2.15a}
    where
    \be
    {\cal A}=\left(\begin{array}{cc}
    0&-1\\
    1&0
    \end{array}\right)\;,
    {\cal A}_1=\left(\begin{array}{cc}
    0&1\\
    1&0
    \end{array}\right)
    \eel{2.15b}

    	The appropriate supermatrix $f^{(0)}_{ij}$ has two eigenvectors
    with zero eigenvalue ($f^{(0)}_{ij}V_j =0$)
    \be
    &&V^{(P)} =\left\{0\;,\;0\;,\;c\;,\ldots\,0\right\} \;, \nonumber \\
    &&V^{(Q)} =\left\{0\;,\;\ldots\;,\,c(x^{-})\;,\;0\;,\;0\right\}
    \eel{2.16}
    where $c$ is a constant, and $c(x^{-})$ is a function.
    This  leads to other two constraints
    \be
    &&\left(\Pi_{\Psi}\Psi\right)^{(P)}=0\;, \label{2.17a} \\
    &&\pa_{-}\Pi^{(Q)}_{-}+ie\left(\Pi_{\Psi}\Psi\right)^{(Q)}=0
    \eel{2.17}
    After the first reduction procedure, the Lagrange density becomes
    \be
    \cL^{(1)}&=&
    \Pi^{(P)} _{-}\dot{A}^{(P)} _{-}+\Pi^{(Q)} _{-}\dot{A}^{(Q)}
    _{-}+\Pi_{\Psi}\dot{\Psi}+{\dot{\lambda}}\left(\pa_{-}\Pi^{(Q)}
    _{-}+ie\left(\Pi_{\Psi}\Psi\right)^{(Q)}\right)\nn \\
    &&-\cH_{F}
    +ie\left(\Pi_{\Psi}\Psi\right)^{(Q)}U_{+}{U}^{-1}_{-}A^{(Q)}_{-}
    -\frac{1}{2}\left(\Pi^{(P)}_{-}\right)^2-\frac{1}{2} \left(
    \Pi^{(Q)}_{-}\right)^{2}
    \eel{2.18}
    where $\lambda$ is a Lagrange multiplier for the constraint (\ref{2.17}).

    The constraint (\ref{2.17a}) is of the first class (in Dirac
classification).
    Therefore one needs a corresponding gauge condition.
    Instead of this, the constraint (\ref{2.17a})
    is to be satisfied on the physical state vector. As a result,
     one can consider the fermionic degrees of freedom, $\Pi_{\Psi}$ and
$\Psi$,
     as independent ones. Using the
    constraints (\ref{2.17}), one can express, in this case, the momentum
    $\Pi^{(Q)}_{-}$ in terms of the fermionic variables
    \be
    \Pi^{(Q)}_{-} =-ie\pa_{-}^{-1}\left(\Pi_{\Psi}\Psi\right)^{(Q)}
    \eel{2.19}
    where $\pa_{-}^{-1}$ is the operator inverse to $\pa_{-}$, and
    whose matrix elements in the coordinate representation (in the  $Q$ sector)
are
    \be
    G^{(Q)}(x^{-}-y^{-})=\frac{\epsilon(x^{-}-y^{-})}{2}-
    \frac{x^{-}-y^{-}}{2L}
    \eel{2.19a}
    (One has to keep in mind that the operator $\pa_{-}^{-1}$ is
    defined only on the $Q$ sector space. Therefore (\ref{2.19})
    represents the only solution to the constraint equation (\ref{2.17}).)

    	The reduced set of variables now is
    \be
    \zeta^{(1)}_{j}&=& \left\{A_{-}^{(P)}\;,\;\Pi^{(P)}_{-}\;,\;
    A_{-}^{(Q)}\;,\;\Pi^{(Q)}_{-}\;,\;\lambda\;,\;\Psi\;,\;\Pi_{\Psi}\right
    \}\;,\;j=1,\ldots,7\;,\nonumber \\
    a^{(1)}_{j}&=&\left\{\zeta^{(1)}_2\;,\;0\;,\;\zeta^{(1)}_4\;,\;0\;,\;
    \pa_{-}\zeta^{(1)}_{4}+ie(\zeta^{(1)}_{7}
    \zeta^{(1)}_{6})^{(Q)} \;,\;\zeta^{(1)}_7\;
    ,\;0\;\right\}
    \eel{2.20}
    The corresponding simplectic supermatrix $f^{(1)}_{ij}$ is still singular,
    but now it has only one eigenvector $V$ with zero eigenvalue
    \be
    V = \left\{0\;,\,0\;,\;-\pa_{-}f(x^-)\;,\;0\;,\;f(x^-)\;,\;-ie\Psi
    f(x^-)\;,\;ie\Pi_{\Psi}f(x^-)\right\}
    \eel{2.21}
    where $f(x^-)$ is a function of $x^-$.
    This vector does not give any new constraints. Consequently,
    according to \cite{FJ}, one needs a gauge condition.

    In the Faddeev-Jackiw quantization procedure, there is only one restriction
    on how one chooses gauge conditions: after imposing gauge conditions, the
    appropriate simplectic supermatrix should be nonsingular.
    On the other hand, gauge conditions should eliminate the gauge freedom.

    Let us consider one of the possibilities of how one
    can choose the gauge condition in the problem considered.
    Using the constraint (\ref{2.17}), the term
    \be
    -ie\frac{U_{+}}{U_{-}}\left(\Pi_{\Psi}\Psi\right)^{(Q)}A_{-}^{(Q)}
    \eel{2.18a}
    in the Lagrange density (\ref{2.18}) can be rewritten as
    \be
    -\frac{U_{+}}{U_{-}}\Pi_{-}^{(Q)}\pa_{-}A_{-}^{(Q)}
    +\frac{U_{+}}{U_{-}}\pa_{-}\left(\Pi^{(Q)}_{-} A_{-}^{(Q)}\right)
    \eel{2.22b}
    A natural gauge condition is chosen to be (see \cite{Brown})
    \be
    \Omega_{G}=\pa_{-}A^{(Q)}_{-}=0
    \eel{2.22}
     From (\ref{2.22}), it
    follows that the nonzero mode $A^{(Q)}_{-}$ is  an arbitrary function of
    $x^+$. On the other hand, the nonzero mode cannot depend only on $x^+$,
    otherwise it is the zero mode by definition. Therefore, the only solution
to the
    eq.(\ref{2.22}) is
    \be
    A ^{(Q)} _{-}=0\;\;({\rm note\; that}\;A_{-}\neq 0)
    \eel{2.22a}
    According to (\ref{2.22}) and the boundary conditions we have chosen,
    the term (\ref{2.22b}) does not give a contribution to the Hamiltonian,
     and can be neglected in the Hamilton density.

    The Lagrange density after the second reduction is
    \be
    \cL^{(2)}&=&\Pi^{(P)}_{-}\dot{A}_{-}^{(P)}+\Pi^{(Q)}_{-}
    \dot{A}^{(Q)}_{-}+\Pi_{\Psi}\dot{\Psi}+
    \dot{\lambda}\left(\pa_{-}\Pi_{-}^{(Q)}+
    ie\left(\Pi_{\Psi}\Psi\right)^{(Q)}\right) \nn \\
    & &+ \dot{\beta }\pa_{-}A^{(Q)}_{-}
    -\cH^{(2)}\;, \nn \\
    \cH^{(2)} &=&\frac{1}{2}(\Pi^{(P)}_{-})^2+\frac{1}{2}(\Pi^{(Q)}_{-})^{2} +
    \cH_{F}
    \eel{2.23}
    where $\beta$ is the Lagrange multiplier for the gauge condition
(\ref{2.22}).
    Now the sets of variables are
    \be
    \zeta^{(2)}_k &=&\left\{A_{-}^{(P)} \;,\;\Pi_{-}^{(P)} \;,\;A_{-}^{(Q)}
    \;,\;\Pi_{-}^{(Q)} \;,\;\la \;,\,\beta \;,\;\Psi\;,\;\Pi_{\Psi}
    \right\}\;,\;k=1,\ldots,8 \;, \nn \\
    a^{(2)}_{k}&=&\left\{\zeta^{(2)}_2\;,\;0\;,\;\zeta^{(2)}_4\;,\;0\;,\;
    \pa_{-}\zeta^{(2)}_4
    +ie(\zeta^{(2)}_8\zeta^{(2)}_7)^{(Q)} \;,\;\pa_{-}\zeta^{(2)}_3\;,\;
    \zeta^{(2)}_8\;,\;0\right\}
    \eel{2.24}
    and the  corresponding simplectic supermatrix has the following
block-diagonal
    form
    \be
    f^{(2)}_{jk}=\left(\begin{array}{cc}
    {\cal A}&{\cal O}\\
    {\cal O}^T&{\cal B}(x^-)\delta(x^- -y^-)
    \end{array}\right)
    \eel{2.25}
    where
    \be
    {\cal B}(x^-)=\left(\begin{array}{cccccc}
    0&-1&0&\pa_{-}&0&0\\
    1&0&\pa_{-}&0&0&0\\
    0&-\pa_-&0&0&ie\Pi_{\Psi}&-ie\Psi\\
    -\pa_-&0&0&0&0&0\\
    0&0&-ie\Pi_{\Psi}&0&0&1\\
    0&0&ie\Psi&0&1&0
    \end{array}\right)
    \eel{2.27}
    the matrix ${\cal A}$ was defined in (\ref{2.15b}),
    the matrix ${\cal O}$ is a zero rectangular $2\times 6$ matrix,
    and the symbol $T$ stands for transposition.

    	The simplectic supermatrix $f^{(2)}_{jk}$ is nonsingular, and can be
    inverted using the Berezin algorithm \cite{Berezin} for the inversion of
    a supermatrix. The result is
    found to be
    \be
    \left(f^{(2)}\right)^{-1}_{jk}=\left(\begin{array}{cc}
    -\frac{1}{2L}{\cal A}&{\cal O}\\
    {\cal O}^T&{\cal B}^{-1}(x^-)\delta(x^- -y^-)
    \end{array}\right)
    \eel{2.28}
    where
    \be
    {\cal B}^{-1}(x^-)=\left(\begin{array}{cccccc}
    0&0&0&-\pa^{-1}_{-}&0&0\\
    0&0&-\pa^{-1}_{-}&0&-ie\pa^{-1}_{-}\Psi&ie
    \pa^{-1}_{-}\Pi_{\Psi}\\
    0&\pa^{-1}_{-}&0&\pa^{-2}_{-}&0&0\\
    \pa^{-1}_{-}&0&-\pa^{-2}_{-}&0&-ie\pa^{-2}_{-}\Psi&
    -ie \pa^{-2}_{-}\Pi_{\Psi}\\
    0&-ie\Psi\pa^{-1}_{-}&0&-ie\Psi\pa^{-2}_{-}&0&1\\
    0&ie\Pi_{\Psi}\pa^{-1}_{-}&0&ie\Pi_{\Psi}\pa^{-2}_{-}&1&0
    \end{array}\right)
    \eel{2.29}
    The operator $\pa_{-}^{-2}$ is the
    one whose matrix elements in the coordinate representation
    (in the $Q$ sector) are
    \be
    H^{(Q)}(x^{-}-y^{-})&=&\frac{|x^{-}-y^{-}|}{2}-\frac{(x^{-}-y^{-})^2}{4L}-
    \frac{2L}{3}
    \eel{2.29a}
    The nonsingularity of the  supermatrix (\ref{2.28}) is consistent with
    the gauge condition (\ref{2.22}) we have chosen.

    	The structure of the supermatrix (\ref{2.28}) shows that the physical
variables are the
    fermionic ones,
    $\Pi_{\Psi}$ and $\Psi$, plus the
    zero modes of $\Pi_- $ and $ A_-$, thereby
    \be
    \omega^{{\rm phys}}=\{A^{(P)} _{-}\;\;,\;\;\Pi^{(P)} _{-}\;\;,\;\;\Psi
    \;\;,\;\;\Pi_{\Psi}\}
    \eel{2.30}
    satisfying the following brackets
    \be
    \left\{A^{(P)} _{-}\;,\;\Pi^{(P)} _{-}\right\}_{FJ}&=&\frac{1}{2L},
    \nn \\
    \left\{\Psi(x^-)\;,\;\Pi_{\Psi}(y^-)\right\}_{FJ}&=&\delta(x^- -y^-)
    \eel{2.31}
    These  brackets coincide with those obtained in  \cite{Brown} using the
    Dirac method of quantization for the systems with ``first-class"
    constraints.

    The quantization procedure consists of the replacement of the variables
$\omega^{{\rm phys}}$ by the corresponding operators
    \be
    \omega^{{\rm phys}}\rightarrow \hat{\omega}^{{\rm phys}}
    \eel{2.31a}
    which obey the following commutation relations
    \be
    &&[\hat{A}_{-}^{(P)},\;\hat{\Pi}_{-}^{(P)}]_{-}=\frac{i}{2L},
    \nn\\
    &&[\hat{\Psi}(x^-),\;\hat{\Pi}_{\Psi}(y^-)]_{+}=i\delta(x^{-} -y^{-})
    \eel{2.31c}

    The Hamilton density corresponding to the physical Hamiltonian can be
    found by solving the constraints (\ref{2.17}), which is equivalent to
    the substitution of $\Pi^{(Q)}_{-}$ (\ref{2.19}) into (\ref{2.23}).
    This gives
    \be
    \cH^{{\rm phys}}=\frac{1}{2}\left(\Pi^{(P)}_{-}\right)^2 +\frac{e^2}{2}
    \left(\pa^{-1}_{-}
    \left(Q\ast \overline{\Psi}\uslash U_- \Psi\right)\right)^2
    +\cH_F
    \eel{2.32}

    It should be mentioned that, in the $2D$ QED with zero modes taken into
    account, there is no real interaction between the gauge field and
    heavy-fermions.

    	Let us discuss the constraint (\ref{2.17a}).
    As it was mentioned above, it is impossible (at least we do not know how to
    do this) to fix the gauge corresponding to this constraint.
    We will consider, for this reason, the constraint
    (\ref{2.17a}) as a strong one,  meaning that this constraint should
    be satisfied on the physical state vectors $|{\rm phys} \hb$
    \be
    :\left(\Pi_{\Psi}\Psi\right)^{(P)}:|{\rm phys}\hb =0
    \eel{2.33}
    where $:...:$ stands for the normal ordering operator.
    The condition (\ref{2.33}) states that the physical state
    vector is chargeless.

    \section{2D Heavy Quark QCD}
    \setcounter{equation}{0}
    \subsection{Quantization of Heavy Quark QCD}
    \vskip2ex

    \hspace{3em} Here we will consider the generalization of the results
    obtained in the
    previous Section to the case of non-Abelian gauge theory, in particular,
    to the two-dimensional heavy-quark QCD.

    We start with the Lagrange density of $2D$ heavy-quark QCD
    \be
    \cL=i\sum_{f}\overline{\Psi}_f \uslash U^{\mu}{\cal D}_\mu \Psi_f-
    M\sum_{f}\overline{\Psi}_{f}\Psi_{f} +\frac12 F^a_{+-}F^a_{+-}
    \eel{3.1}
    where the index $f$ stands for the flavor of the quark, $\Psi_{f}$
    is a color multiplet of quarks with a given flavor $f$, and it is  assumed
    that the quark's mass is flavor independent,
    \be
    {\cal D}_\mu=\pa_\mu +igA^a_\mu T^a
    \eel{3.2}
    is the covariant derivative, $\;T^a\;\;(a=1,\ldots,N^{2}_{c}-1)$
    are the generators of Lie algebra corresponding to the group SU(${\rm
N}_c$),
    obeying the following commutation relations
    \be
    \left[T^a\;,\;T^b\right]_{-}=if^{abc}T^c
    \eel{3.3}
    with the structure constants $f^{abc}$ being antisymmetric in all indices,
    and $F^a_{+-}$ is the field strength tensor
    \be
    F^a_{+-}&=&\pa_+A^a_- -\pa_-A^a_+ +g\left(A_+\times A_-\right)^a
    \;\;,\;\; \\
    A^a_{\pm}&=&\frac{1}{\sqrt{2}}(A^a_0 \pm A^a_3)
    \eel{3.4}
    Here, for any two ``isotopic" vectors, say, $\cF^a $ and $\cG^b$,
    the cross product is defined as
    \be
    (\cF \times \cG )^a =f^{abc}\cF^b \cG^c
    \eel{3.4a}
    For simplicity, in what follows, we will consider flavorless quarks.

    	The canonical momenta to the variables under consideration are
    \be
    &&\Pi_{\Psi}=i \overline {\Psi}\uslash
U_-\;\;,\;\;\Pi_{\overline{\Psi}}=0\;,\nonumber \\
    &&\Pi^{a (P)}_- =P\ast \frac{\pa \cL}{\pa \dot{A}_-}=
    \dot{A}^{a(P)} _- +g\left(A_+\times A_-\right)^{a(P)} \;, \nn \\
    &&\Pi^{a (Q)}_- =Q\ast \frac{\pa \cL}{\pa \dot{A}_-}=
    \dot{A}^{a (Q)} _- -\pa _-A ^{a(Q)} _+
    +g\left(A_+\times A_-\right)^{a(Q)} \;, \nn \\
    &&\Pi ^{a(P,Q)}_+ =0
    \eel{3.5}
    The  velocities $\dot{A}^{a (P,Q)}_{-}$ can be expressed
    through the momenta
    $\Pi^{a (P,Q)}_{-}$, and one has the following  primary constraints
    ${\bf \Phi}= 0$
    \begin{equation}
    {\bf \Phi}=\left\{ \begin{array}{ll}
    		 \phi^{a (P,Q)}_{+}=\Pi^{a(P,Q)}_{+}\\
    		 \phi_{\Psi}=\Pi_{\Psi}-i\overline{\Psi}
    		 \uslash {U}_{-}\\
    		 \phi_{\overline{\Psi}}=\Pi_{\overline{\Psi}}
    		 \end{array}
    	 \right.
    \label{3.5a}
    \end{equation}

    	On the surface of these constraints, the Lagrange density
    (\ref{3.1}) in the  Hamilton form reads as
    \be
    \cL = \Pi^{a (P)}_{-} \dot{A}^{a (P)}_{-}+\Pi^{a (Q)}_{-}
    \dot{A}^{a (Q)}_{-}+\Pi_{\Psi}\dot{\Psi}-\cH  \;,
    \eel{3.6}
    where
    \be
    \cH&=& \cH_{F}+\cH_{G} , \nn \\
    \cH_{F}&=&-\frac {U_+}{U_-} \Pi_{\Psi}\pa _-
    \Psi -
    i\frac {M}{U_-} \Pi_{\Psi} \uslash \Psi ,\nn \\
    \cH_{G}&=&\frac{1}{2}\left(\Pi^{a (P) 2}_{-}+\Pi^{a (Q) 2}_{-}\right)
    +\Pi^{a (Q)}_{-}\pa_-A^{a (Q)}_+
    -g \Pi^{a (P)}_{-} \left(A_+\times A_-\right)^{a (P)}\nn \\
    &&-
    g \Pi^{a (Q)}_{-} \left(A_+\times A_-\right)^{a (Q)}
    -ig \frac{U_+}{U_-} \Pi_{\Psi} T^a \Psi
    \left(A^{a (P)}_{-} +A^{a (Q)}_{-}\right)\nn \\
    &&-ig \Pi_{\Psi} T^a \Psi
    \left(A^{a (P)}_{+} +A^{a (Q)}_{+}\right)
    \eel{3.7}
    Using the similar arguments from the previous Section, we neglect the term
    $\Pi^{a (P)}_{-}\Pi^{a (Q)}_{-}$ in the Hamilton density $\cH _{G}$.

    	The sets of the initial variables
    $\zeta^{(0)}$ and the ``generalized'' momenta $a^{(0)}(\zeta)$ are
    \be
    &&\zeta^{(0)}_{j} = \left\{A_{-}^{a (P)}\;,\;\Pi^{a (P)}_{-}\;,\;
    A^{a (P)}_{+}\;,\;
    A_{-}^{a (Q)}\;,\;\Pi^{a (Q)}_{-}\;,\;A^{a (Q)}_{+}\;,\;\Psi\;,\;
    \Pi_{\Psi}\right \}\;, \nn \\
    && a^{(0)}_{j}=\left\{\zeta^{(0)}_2\;,\;0\;,\;0\;,\;\zeta^{(0)}_5\;,\;
    0\;,\;0\;,\;\zeta^{(0)}_8\;
    ,\;0\;\right\}\;\;,\;\;j=1,\ldots,8
    \eel{3.8}
    The corresponding simplectic supermatrix $f^{(0)}_{ij}$  has two
eigenvectors
    with zero eigenvalues
    \be
    &&V^{(P)} =\left\{0\;,\;0\;,\;c^a\;,\ldots\,0\right\} \;, \nn \\
    &&V^{(Q)} =\left\{0\;,\;\ldots\;,\,c^{a}(x^{-})\;,\;0\;,\;0\right\}
    \eel{3.9}
    leading to two other sets of constraints
    \be
    &&ig\left( \Pi_{\Psi}T^a \Psi\right)^{(P)} +g\left(A_{-}\times \Pi_{-}
    \right) ^{a (P)}=0\;,
    \label{3.10}\\
    &&\pa _{-}\Pi^{a (Q)}_{-}+g\left(A_{-}^{(P)} \times \Pi_{-}^{(Q)}
    \right)^a +g\left(A_{-}^{(Q)} \times \Pi_{-}
    \right)^a
    +ig\Pi_{\Psi}T^a \Psi=0
    \eel{3.11}
    Equation (\ref{3.10}) tells that the total classical color charge in the
    system is zero.

    The Hamilton density $\cH_G$ on the constraint surface
    (\ref{3.5a}), (\ref{3.10}), and (\ref{3.11}) becomes
    \be
    \cH_G =\frac{1}{2}\left(\Pi^{a (P) }_{-}\right )^2+\halv \left(\Pi^{a (Q) }
    _{-}\right)^2 -ig\frac{U_+}{U_-}\Pi_{\Psi}T^a \Psi
    ( A^{a (P)}_{-}+A^{a (Q)}_{-} )
    \eel{3.12}
    If one adds  the constraints
    (\ref{3.10}) and (\ref{3.11}) with corresponding Lagrange multipliers
    to the Lagrange density (\ref{3.6}), this does not
    give any new constraints. Therefore, one needs gauge conditions.
    Let us choose them, like in QED,  in the form
    \be
    \Omega_{G}^{a} =\pa _{-}A^{a (Q)}_{-} =0
    \eel{3.13}
    Using the same arguments as in the previous Section, one obtains
    \be
    A^{a (Q)}_{-} = 0,\;\;\left({\rm note\; that}\;A^{a}_{-}\neq 0\right)
    \eel{3.13a}
    The Hamilton density $\cH_G$ then can be sufficiently simplified
    \begin{eqnarray*}
    \cH_G &=&\frac{1}{2}\left(\Pi^{a (P) }_{-}\right )^2
    +\halv \left(\Pi^{a (Q) }_{-}\right)^2
    -ig\frac{U_+}{U_-}\Pi_{\Psi}T^a \Psi
     A^{a (P)}_{-}
    \end{eqnarray*}
    It can be shown, using the constraints (\ref{3.10}), that the last
    term in the above expression does not give a contribution to the
    Hamiltonian $H_G$.
    Keeping this in mind, one can neglect it in the Hamilton density.
    Therefore, one can write
    \be
    \cH_G &=&\frac{1}{2}\left(\Pi^{a (P) }_{-}\right )^2
    +\halv \left(\Pi^{a (Q) }_{-}\right)^2
    \eel{3.14}
    where the variables $\Pi^{a (Q)}_{-} $ have to be eliminated
    using the Gauss law constraints
    \be
    \Omega^{a}_{GL}=\pa_{-}\Pi^{a (Q)}_{-}+g\left(A_{-}^{(P)} \times
\Pi_{-}^{(Q)}
    \right)^a -g\left(A_{-}^{(P)}\times\Pi_{-}^{(P)}\right)^a +ig\rho^{a}=0
    \eel{3.14a}
    where
    \be
    \rho^{a}= \left(\Pi_{\Psi}T^a \Psi\right)^{(Q)}
    \eel{3.15}
    In deriving (\ref{3.14a}), we have used the relation
    \begin{eqnarray*}
    (A_{-} \times \Pi_{-})^{a (P)}=(A^{(P)}_- \times \Pi^{(P)}_{-})^a
    \end{eqnarray*}
    which holds on the surface of the gauge condition
    (\ref{3.13}) (or (\ref{3.13a})).

    Now the corresponding simplectic supermatrix $f^{(2)}_{ij}$ is nonsingular,
     and has a similar form as in 2D QED with
    \be
    \{A^{a (P)}_{-} \;,\;\Pi^{b (P)}_{-}\}_{FJ}&=&\frac{1}{2L}\delta^{ab}
    \;,\nonumber \\
    \{\Pi_{\Psi}(x^-)\;,\;\Psi(y^-)\}_{FJ}&=&\delta(x^-
    -y^-)
    \eel{3.15a}
    Therefore, one can consider the set of the variables
    \be
    \omega^{{\rm phys}}=\left\{(A_{-}^{a(P)},\;\Pi^{a (P)}_{-},\;
    \Pi_\Psi,\;\Psi \right\}
    \eel{3.15b}
    as physical ones.
    The quantization procedure consists of the replacement of the variables
    $\omega^{{\rm phys}}$ by the corresponding operators
    \be
    \omega^{{\rm phys}}\rightarrow \hat{\omega}^{{\rm phys}}
    \eel{3.15c}
    which obey the following commutation and anticommutation relations
    \be
    &&[\hat{A}_{-}^{a (P)},\;\hat{\Pi}_{-}^{b
(P)}]_{-}=\frac{i}{2L}\delta^{ab},
    \label{3.15d}\\
    &&[\hat{\Psi}(x^-),\;\hat{\Pi}_{\Psi}(y^-)]_{+}=i\delta(x^- -y^-)
    \eel{3.15e}

    \subsection{Physical Hamiltonian}

    \hspace{3em}To obtain the physical Hamiltonian, one has to solve the
constraints
    (\ref{3.14a}) in order to express the momenta $\Pi^{a (Q)}_{-}$ in terms of
    physical variables.
    Although the eq.(\ref{3.14a}) is the one for the nonzero mode of the
momenta
    $\Pi ^{a (Q)}_{-}$, its left-hand-side contains a term which is a zero
    mode (the product of the zero modes does not have the nonzero mode).
    Thus, one has to be careful when such an equation has to be solved.

    	Let us define the function
    \be
    Z^a (x^-) =V^{ab}(x^-)\Pi ^{a (Q)}_{-}(x^-)
    \eel{a1}
    where
    \be
    V^{ab}(x^-)=\left(\exp (gRx^- )\right)^{ab},\;\;\left(
    R^{ab}=f^{acb}A^{c (P)}_{-}\right)
    \eel{a2}
    is a unitary matrix. The Hamilton density (\ref{3.14}) can be expressed
    through the function $Z^a (x^-)$
    \be
    \cH_G &=&\frac{1}{2}\left(\Pi^{a (P) }_{-}\right )^2
    +\halv \left(Z^a (x^-)\right)^2
    \eel{a3}
     From (\ref{3.14a}),  the equation that the function $Z^a (x^-)$
    satisfies is found to be
    \be
    \pa _{-}Z^a (x^-) &=&\left(V(x^-)L(x^-)\right)^a ,\label{3.21}\\
    L^{a}(x^-)&=&
    g\left(A_{-}^{(P)} \times \Pi_{-}^{(P)}
    \right)^a -ig\rho^{a} (x^-)
    \eel{a4}

    The solution to the eq.(\ref{3.21}) has the form
    \be
    Z^{a}(x^-)= \left(VL\right)^{a (P)}x^{-}+
    \pa_{-}^{-1} \left(VL\right)^{a (Q)}(x^-)
    \eel{3.22}

    	We wish to stress out the existence of the linear term in
    (\ref{3.22}). As we will see later, this term will give a contribution
    to the physical Hamiltonian where the momenta $\Pi^{a (P)}_{-}$ are
    involved.

    	To obtain the zero modes and nonzero modes of the expressions
    involved in (\ref{3.22}), we remark that
    \be
    L^{a (P)}&=&g\left(A_{-}^{(P)} \times \Pi_{-}^{(P)}\right)^a\;\;,\;\;
    L^{a (Q)}=-ig\rho^{a}
    \eel{3.23}
    Substituting these expressions into (\ref{a1}), one obtains
    the function $Z^{a }(x^{-}) $ in terms of the physical variables
$\omega^{{\rm
    phys}}$ defined by (\ref{3.15b})
    \be
    &&Z^{a }(x^{-}) =g\left\{V^{ac (P)}
    \left(A ^{(P)}_-  \times \Pi ^{(P)}
    _-\right)^c-iK^{a (P)}\right\}x^{-} \nn \\
    &&+g\pa _{-}^{-1}\left\{(V^{ac (Q)}(x^-)
    \left( A ^{(P)}_-  \times \Pi ^{(P)} _-\right)^c
    -iK^{b (Q)} -iV^{bc (P)}\rho^{c}(x^-)\right\}
    \eel{3.24}
    where
    \be
    K^a (x^-)=V^{ab (Q)}(x^-) \rho^{b}(x^-)
    \eel{3.25}

    	The eq.(\ref{3.24}) makes possible to construct the physical
    Hamiltonian $H^{{\rm phys}}$
    \be
    H^{{\rm phys}}&=&\int_{-L}^{L} \cH^{{\rm phys}}dx^{-}\;, \\
    \label{3.26}
    \cH^{{\rm phys}}&=&\cH_{F}+\halv (\Pi^{a (P)}_{-} )^2 +
    \halv (Z^{a }(x^-) )^2
    \eel{3.27}
    where  $Z^{a }(x^{-}) $ is given by (\ref{3.24})
    and  can be presented in the following form
    \be
    Z^{a}(x^-) &=&gB^{a (P)} x^{-}+g \pa _{-}^{-1}B ^{a (Q)}(x^{-})\;,\\
    \label{3.28}
    B^{a (P)}&=&G^{a (P)} -iK^{a (P)}\;, \nn \\
    B^{a (Q)}(x^-)&=&G^{a (Q)}(x^-)
     -i{\cal P}^{a (Q)}(x^-) \nn \\
    G^{a}(x^-)&=&V^{ab}(x^-)\left(A ^{(P)}_-  \times \Pi ^{(P)}_-\right)^{b}
    =V^{ab}(x^-)R^{bc}\Pi^{c (P)}_{-} \;, \nn \\
    {\cal P}^{a (Q)}(x^-)&=&K^{a (Q)}(x^-)+
    V^{ab (P)} \rho^b
    \eel{3.29b}

    	Establishing the necessary tools, we can now consider the
    interaction Hamiltonian.
    Due to eqs.(\ref{3.27}) and (\ref{3.28}), the interaction
    Hamiltonian corresponds to the last term in (\ref{3.27})
    \be
    H_{{\rm int}}=\frac{g^2}{2}\int_{-L}^{L} \cH_{{\rm int}}(x)dx
    \eel{3.301}
    where
    \be
    \cH_{{\rm int}}(x)=B^{a (P)} B^{a (P)} x^2 +2xB^{a (P)}
    \pa^{-1}B ^{a (Q)}(x) +\pa ^{-1}B ^{a(Q)}(x)\pa ^{-1}B ^{a (Q)}(x)
    \eel{3.302}

    	One can present the interaction Hamiltonian in the following form
    \be
    H_{{\rm int}}=H_{{\rm int}}^{{\rm zero}}+H_{{\rm int}}^{{\rm linear}}+
    H_{{\rm int}}^{{\rm quad}}
    \eel{3.303}
    where $H_{{\rm int}}^{{\rm zero}}$ is the part of interaction
    Hamiltonian which describes the self-interaction of the gauge zero modes
    in the sector of the pure Yang-Mills fields
    \be
    H_{{\rm int}}^{{\rm zero}}&=&\frac{g^2}{2}\left[\frac{2L^3}{3}G^{a (P)}
    G^{a (P)}+\int_{-L}^{L}\pa ^{-1}G^{a (Q)}(x)\pa ^{-1}G^{a (Q)}(x)
    dx \right. \nonumber \\
    &&\left. -2G^{a (P)}\int_{-L}^{L}(\frac{x^2}{2}-\frac{L^2}{6})
    G^{a (Q)}(x)dx
    \right]
    \eel{3.304}
    the term $H_{{\rm int}}^{{\rm linear}}$ is the part of interaction
    Hamiltonian which is linear over the quark color charge $\rho^{a}$
    \be
    H_{{\rm int}}^{{\rm linear}}&=&-ig^2\left[\frac{2L^3}{3}G^{a (P)}K^{a
    (P)}+\int_{-L}^{L}\pa ^{-1}G^{a (Q)}(x)\pa ^{-1}\cP^{a (Q)}(x)
    dx \right. \nn \\
    && \left. -\int_{-L}^{L} (\frac{x^2}{2}-\frac{L^2}{6})
    \left(G^{a (P)} \cP^{a (Q)}(x)+
    K^{a (P)} G^{a (Q)}(x)\right)dx \right]
    \eel{3.30b}
    and the term $H_{{\rm int}}^{{\rm quad}}$ describes the
    quark-quark interaction
    \be
    H_{{\rm int}}^{{\rm quad}}&=&-\frac{g^2}{2}\left[\frac{2L^3}{3}
    K^{a (P)}K^{a (P)}+\int_{-L}^{L}\pa ^{-1}\cP^{a (Q)}(x)\pa ^{-1}
    \cP^{a (Q)}(x)dx \right. \nn \\
    &&\left.- 2K^{a (P)}\int_{-L}^{L} (\frac{x^2}{2}-
    \frac{L^2}{6})\cP^{a (Q)}(x)dx \right]
    \eel{3.30c}
    In deriving  these expressions, we used the following properties of the
    operator $\pa^{-1}$
    \be
    \int_{-L}^{L}(\pa ^{-1}f_{1}^{(Q)})(x)f_{2}^{(Q)}(x)d x &=&
    -\int_{-L}^{L}(\pa ^{-1}f_{2}^{(Q)})(x)f_{1}^{(Q)}(x)d x\;,
    \nn \\
    \pa^{-1}x&=&\frac{x^2}{2}-\frac{L^2}{6}
    \eel{3.30d}

    For the Abelian case, the only term that
    survives in (\ref{3.303})
    is $H_{{\rm int}}^{{\rm quad}}$, given by (\ref{3.30c}) with
    $K^{a }=0$ and $\cP^{(Q)}(x)=\left(\Pi_{\Psi}\Psi\right)^{(Q)}$. This gives
    the result obtained in the previous Section.

    Let us now return to the constraint (\ref{3.10}). Using the gauge condition
    (\ref{3.13}) (or (\ref{3.13a})), it can be rewritten as
    \be
    i\left( \Pi_{\Psi}T^a \Psi\right)^{(P)} +\left(A_{-}^{(P)}\times
    \Pi_{-}^{(P)}\right) ^{a}=0
    \eel{3.30f}
    This constraint is of the
    first class.  We do not know how to fix the gauge corresponding to this
    constraint. Therefore, as in the previous Section, we will consider the
    constraint
    (\ref{3.30f}) as a strong one, meaning that it should be satisfied on the
    physical state vectors $|{\rm phys}\hb$
    \be
    :\left(i\left(\Pi_{\Psi}T^a \Psi\right)^{(P)}+\left(A_{-}^{(P)}\times
    \Pi_{-}^{(P)}\right) ^{a}\right):|{\rm phys}\hb =0
    \eel{3.30g}
    where $:...:$ stands for the normal ordering operator.

    \subsection{The group SU(2)}

    \hspace{3em}Let us examine the results obtained for the SU(2)
    group. In this case, the structure constants $f^{abc}$ coincide with the
    antisymmetric Levi-Civita tensor $\epsilon^{abc}$
    \be
    f^{abc}=\epsilon^{abc}\;\;,\;\; a=1,2,3,\;\;\;(\epsilon^{123}=1)
    \eel{3.31}

    The matrix  $V^{ab}(x^-)$ can be calculated and is found to be
    \be
    V^{ab}(x^-)&=&\delta^{ab}+V_1(x^-)\frac{(R^2)^{ab}}{D^2}+V_2(x^-)
    \frac{R^{ab}}{D}\;, \nn\\
    V_1(x^-)&=&1-\cos\left(gx^{-}D\right)\;\;,\;\;V_2(x^-)=
    \sin\left(gx^{-}D\right)
    \eel{3.33}
    where
    \be
    D=\sqrt{A^{a (P)}_{-} A^{a (P)}_{-}}\;\;,\;\;\left(R^{2}\right)^{ab}
    =A^{a (P)}_{-} A^{b (P)}_{-}-\delta^{ab}D^2
    \eel{3.34}

    	From these expressions, one can explicitly find the zero modes and
    nonzero modes of the matrix $V^{ab}(x^-)$ involved in the physical
Hamiltonian
    \be
    V^{ab
(Q)}(x^-)&=&u_1(x^-)\frac{(R^2)^{ab}}{D^2}+u_2(x^-)\frac{R^{ab}}{D}\;,
    \nn \\
    V^{ab (P)}&=&\delta^{ab}-u_1(0)\frac{(R^2)^{ab}}{D^2}\;,\nn \\
u_{1}(x^-)&=&\frac{\sin(gDL)}{gDL}-\cos(gDx^-)\;\;,\;\;u_{2}(x^-)=\sin(gDx^-)
    \eel{3.35}
    Other quantities that are involved in the problem are
    \be
    \pa_{-}^{-1}V^{ab (Q)}(x^-)&=&v_1(x^-)\frac{(R^2)^{ab}}{D^2}+
    v_2(x^-)\frac{R^{ab}}{D}\;,\nn \\
    v_1(x^-)&=&-\frac{1}{gD}u_{2}(x^-)+x^{-}
    \frac{\sin(gDL)}{gDL}\;\;,\;\;v_2(x^-)=\frac{1}{gD}u_1(x^-)\;,
    \label{3.36} \\
    \pa_{-}^{-2}V^{ab (Q)}(x^-)&=&z_1(x^-)\frac{(R^2)^{ab}}{D^2}+
    z_2(x^-)\frac{R^{ab}}{D}\;, \nn \\
    z_1(x^-)&=&-\frac{1}{g^{2}D^{2}}u_{1}(x^-)+
    (\frac{(x^-)^2}{2}-\frac{L^2}{6})
    \frac{\sin(gDL)}{gDL}\;,\nn \\
    z_2(x^-)&=&
    -\frac{1}{g^{2}D^{2}}u_{2}(x^-)+\frac{x^-}{gD}\frac{\sin(gDL)}{gDL} \;,
    \label{3.37}\\
    G^{a (P)}&=&\frac{\sin gDL}{gDL}R^{ab}\Pi^{b (P)}_{-}\;, \nn \\
    G^{a (Q)}(x^{-})&=-&D\left(u_1(x^{-})\frac{R^{ab}}{D}-u_2(x^{-})
    \frac{(R^2)^{ab}}{D^2}
    \right)\Pi^{b (P)}_{-}\;, \nn \\
K^{a}(x^{-})&=&\left(u_1(x^{-})\frac{(R^2)^{ab}}{D^2}
+u_2(x^{-})\frac{R^{ab}}{D}
    \right)\rho^{b}(x^{-})
    \eel{3.38}

    	The part $H_{{\rm int}}^{{\rm zero}}$ of the interaction Hamiltonian
    can be simplified and presented in the following form
    \be
    &&H_{{\rm int}}^{{\rm zero}}
    =-L\left[ 1-\left(\frac{\sin(gDL)}{gDL}\right)^2 \right]
    \Pi^{a (P)}_{-} \frac{(R^2)^{ab}}{D^2}\Pi^{b (P)}_{-}
    \eel{3.39}
    This term describes the self-interaction of the gluons in terms of
    their zero modes.

    	The result for the part of the interaction Hamiltonian
    $H_{{\rm int}}^{{\rm linear}}$ is
    \be
    H_{{\rm int}}^{{\rm linear}}=-ig^2 D\left( \int _{-L}^{L} \la_1 (x)
    \Pi^{a (P)}_{-} \frac{(R^2)^{ab}}{D^2}\rho^b (x)dx +
    \int _{-L}^{L} \la_2 (x)
    \Pi^{a (P)}_{-} \frac{R^{ab}}{D}\rho^b (x)dx\right)
    \eel{3.40}
    where
    \be
    \la_1 (x)&=&-\frac{x}{gD}\frac{\sin(gDL)}{gDL}\cos(gDx)+
    \frac{\cos(gDL)}{g^2 D^2}\sin(gDx) , \nn \\
    \la_2 (x)&=&\frac{x}{gD}\frac{\sin(gDL)}{gDL}\sin(gDx)-
    \frac{1}{g^2 D^2}\cos(gDL)\left(\frac{\sin(gDL)}{gDL}-\cos(gDx)\right)
    \eel{3.41}
    This term describes the interaction of the heavy-quarks with the zero modes
    of the gauge degrees of freedom.

    	The part $ H^{{\rm quad}}_{{\rm int}}$
    of the interaction
    Hamiltonian, which describes the self-interaction between heavy-quarks,
     has the following form
    \be
    &&H_{{\rm int}}^{{\rm quad}} =g^2\int_{-L}^{L} \int_{-L}^{L}
    \left(A_1 (x,y)\rho^a (x)\frac{(R^2)^{ab}}{D^2}\rho^b (y)+
    A_2 (x,y)\rho^a (x)\frac{R^{ab}}{D}\rho^b (y)\right)dx dy \nn\\
    &&+g^2 \int_{-L}^{L} \int_{-L}^{L}\rho^a(x)\left(\delta^{ab}+
    \frac{(R^2)^{ab}}{D^2}\right)\rho^b (y)H^{(Q)}(x-y)dx dy
    \eel{3.42}
    where
    \be
    A_1 (x,y)&=&\frac{L}{6}\left(u_1(x)u_1(y) +u_2(x)u_2(y)\right)\nn \\
    &&+\frac{1}{2L}\left\{\left[(\frac{x^2}{2}-\frac{L^2}{6})\left(u_1(x)
    \cos(gDy)
    -u_2(x)\sin(gDy)\right)\right]+\left[x\leftrightarrow
    y\right]\right\}\nn \\
    &&+\halv \left[\left(u_1(x)\cos(gDy)
    -u_2(x)\sin(gDy)\right)+\left(x\leftrightarrow
    y\right)\right]H^{(Q)}(x-y)\nn\\
    &&-\frac{\sin(gDL)}{2gDL}
    \left(\cos(gDx)+\cos(gDy)\right)H^{(Q)}(x-y),
    \label{3.43}\\
    A_2(x,y)&=&\frac{L}{6}\left(u_1(x)u_2(y) -u_2(x)u_1(y)\right)\nn \\
    &&-\frac{1}{2L}\left\{\left[(\frac{x^2}{2}-\frac{L^2}{6})\left(u_2(x)
    \cos(gDy)
    +u_1(x)\sin(gDy)\right)\right]-\left[x\leftrightarrow
    y\right]\right\}\nn \\
    &&-\halv \left[\left(u_2(x)\cos(gDy)
    +u_1(x)\sin(gDy)\right)-\left(x\leftrightarrow
    y\right)\right]H^{(Q)}(x-y)\nn\\
    &&+\frac{\sin(gDL)}{2gDL}
    \left(\sin(gDx)-\sin(gDy)\right)H^{(Q)}(x-y)
    \eel{3.44}
    where $H^{(Q)}(x-y)$ is given by (\ref{2.29a}).

    	From (\ref{3.27}) and (\ref{3.39}), one can find the total
    Hamiltonian corresponding to the pure Yang-Mills fields
    \be
    H_{{\rm pure\;YM}}=L\Pi^{a (P)}_{-}G^{ab}\Pi^{b (P)}_{-}
    \eel{3.45}
    where
    \be
    G^{ab}=\delta^{ab}-\left(1-\left(\frac{\sin gDL}{gDL}\right)^2 \right)
    \frac{(R^2)^{ab}}{D^2}
    \eel{3.46}
    The Hamiltonian $H_{{\rm pure\;YM}}$ is quadratic in the momenta
    $\Pi^{a (P)}_{-}$ and sufficiently nonlinear in the coordinates
    $A^{a (P)}_{-}$.

    \subsection{The Faddeev-Popov Determinant}

    \hspace{3em}The constraints (\ref{3.14a}) and gauge conditions
    (\ref{3.13}) are the only
    local constraints in the problem. Although these constraints are
    linear over the variables $\Pi^{a (Q)}_{-} $ and $A^{a (Q)}_{-} $, their
    Poisson bracket is nontrivial and depends on the physical variables
    $A^{a (P)}_{-} $. This might give a nonzero contribution to the effective
    interaction potential. To find such a contribution, we should calculate
    the Faddeev-Popov determinant
    which appears due to the constraints (\ref{3.13}) and (\ref{3.14a}). One
    should introduce the Faddeev-Popov ghosts provided that
    the Faddeev-Popov determinant is not trivial (it may depend
    on the field variables). On the other hand, the Faddeev-Popov
    determinant is important for the unitarity of the S-matrix.

    In order to find the Faddeev-Popov determinant,  one should consider the
    Poisson bracket between the first-class constraints (\ref{3.14a}) and the
    corresponding gauge conditions (\ref{3.13}). The result is found to be
    \be
    \{\Omega^{a}_{G}(x)\;,\;\Omega^{b}_{GL}(y)\}_{PB}&=&-\pa_{x}
    {\cal D}^{ab}_{-}(x)D^{(Q)}(x -y)\;,\\
    \label{f1}
    {\cal D}^{ab}_{-}(x)&=&\delta^{ab}\pa_{x}+gR^{ab},\;\;R^{ab}
    =f^{acb}A^{c (P)}_{-}
    \eel{f2}
    where
    \be
    D^{(Q)}(x -y)=\delta(x -y)-\frac{1}{2L}
    \eel{f3}
    is the $\delta$-function in the $Q$ sector. (For simplicity, we
    omitted the index ``-" in the coordinates $x$ and $y$.)

    The Faddeev-Popov determinant is defined as
    \be
    \Delta[A]={\rm Det}^{1/2}\{\Psi_\ell ,\Psi_{\ell'}\}_{PB}|_{\Psi
    =0}
    \eel{f3a}
    where $\Psi_\ell$ is the  set of first class constraints and the
corresponding gauge conditions.
    In our case, the Faddeev-Popov determinant  depends only on the variables $
    A^{(P)}_{-}$ and can be presented as
    \be
    \Delta [A^{(P)}_{-}] &=&{\rm Det}\left(\frac{
    \pa_- {\cal D}_{-}}{\pa ^{2}_{-}}\right)={\rm Det}\left(\frac{
    {\cal D}_{-}}{\pa_{-} }\right)
    \eel{f3b}
    In order to normalize the Faddeev-Popov determinant, we have introduced the
operator $\pa^{2}_{-}$ in the denominator of (\ref{f3b}).
    Using the well known property of the determinant of a matrix
    \begin{eqnarray*}
    &&{\rm Det}M =\exp \left({\rm Tr\; Ln}M \right)
    \end{eqnarray*}
    one obtains
    \be
    \Delta [A^{(P)}_{-}]= \exp \left({\rm Tr\; Ln}\frac{{\cal D}_{-}}
    {\pa_{-} }\right)
    \eel{f4}

    Taking the derivative of the
    both sides of (\ref{f4}) with respect to the coupling constant $g$,  one
    obtains an ordinary differential
    equation for the Faddeev-Popov determinant
    \be
    \frac{d\Delta[A^{a (P)}_-]}{dg}&=&\left({\rm Tr}{\cal D}_{-}^{-1}(g)
    R\right)\Delta[A^{a (P)}_-]
    =\left({\rm Tr}G(|g)R \right)\Delta[A^{a (P)}_-]
    \eel{f13}
    where $G^{ab}(x,y|g)$ is the Green function of the operator ${\cal
    D}_{-}^{ab}$.
    The solution of this equation has the following form
    \be
    \Delta[A^{a (P)}_-]=\exp\left(\int^{g}_{0}dg'{\rm Tr}G(|g')R
    \right)
    \eel{f15}

    The determinant (\ref{f3b}) is equivalent to the Faddeev-Popov term in
    the effective action
    \be
    S_{FP}=\int dx^+ dx^- \bar{c}^{a}(x)\pa_{-}\cD ^{ab}_{-}(x)c^{b}(x)
    \eel{f15a}
    where $c^b$ and $\bar{c}^a$ are the ghost  and antighost fields,
    respectively. The operator
    $\cD ^{ab}_{-}(x)$ depends only on the zero modes $A_{-}^{a (P)}$.
    Therefore, only the nonzero modes of the ghost and antighost fields give
    contributions to the Faddeev-Popov action, meaning that the
     Faddeev-Popov ghosts are necessary  only in their own $Q$-sector.
    One should then make the following substitution
    \begin{eqnarray*}
    &&\bar{c}^{a} \ra \bar{c}^{a (Q)},\;\; c^{a} \ra  c^{a (Q)}
    \end{eqnarray*}
     and the Faddeev-Popov action becomes
    \be
    S_{FP}=\int dx^+ dx^- \bar{c}^{a(Q)}(x)\pa_{-}
    \cD ^{ab}_{-}(x)c^{b(Q)}(x)
    \eel{f15b}
    where $\bar{c}^{a(Q)}$ and $c^{a(Q)}$ are the nonzero modes of the
    Faddeev-Popov ghosts.

    	Let us compute the Faddeev-Popov determinant. As it follows from
    (\ref{f15}), one needs to know the Green
     function $G^{ab}(x,y|g)$. According to (\ref{f15b}), it satisfies the
    equation
    \be
    \pa_x G^{ab}(x,y|g)+gR^{ac}G^{cb}(x,y|g)=\delta^{ab}D^{(Q)}(x-y)
    \eel{f15aa}
    The matrix $R^{ab}$ does not depend on $x$. Consequently, the Green
function
    $G^{ab}(x,y|g)$ is translational invariant:
    $G^{ab}(x,y|g)=G^{ab}(x-y|g) =G^{ab}(z|g)$, and (\ref{f15a}) can be
rewritten as
    \be
    \pa^{z}_{-}G^{ab}(z|g)+gR^{ac}G^{cb}(z|g)=\delta^{ab}D^{(Q)}(z)
    \eel{f16}
    This equation is an inhomogeneous first order differential equation, and
    its solution can be presented in the following form
    \be
    G^{ab}(z|g)=U^{ac}(z|g)\left(V^{cb}(|g)D^{(Q)}\right)^{(P)}z
    +U^{ac}(z|g)\pa^{-1}\left(V^{cb}(|g)D^{(Q)}\right)^{(Q)}(z)
    \eel{f17}
    where the matrix $V^{ab}(z|g)$ was defined by (\ref{a2}), and the matrix
    $U^{ab}(z|g)$ is its inverse one.
    Substituting (\ref{f17}) into (\ref{f15}), one obtains
    \be
    \Delta[A^{a (P)}_-]=\exp\left(2L\int^{g}_{0}dg'
    \pa^{-1}\left(V^{ab}(|g')D^{(Q)}\right)^{(Q)}(z)|_{z=0} R^{ba}\right)
    \eel{f18}
    The integrand in the right-hand-side of eq. (\ref{f18}) can be presented
    as
    \be
    &&\pa^{-1}\left(V^{ab}(|g')D^{(Q)}\right)^{(Q)}(z)=\nn\\
    && =V^{ab (P)}(g')G^{(Q)}(z)+
    \pa^{-1}\left(V^{ab (Q)}(|g')D^{(Q)}\right)^{(Q)}(z)
    \eel{f19}
    where the function $G^{(Q)}(z)=\frac{\epsilon (z)}{2}-\frac{z}{2L}$
    was defined by (\ref{2.19a}). Thus
    \be
    &&\Delta[A^{a (P)}_-]=\exp\left(2L\int^{g}_{0}dg'
    \pa^{-1}\left(V^{ab (Q)}(|g')D^{(Q)}\right)^{(Q)}(z)|_{z=0}
    R^{ba}\right)
    \eel{f21}

    	This is, so far, the most general result we can obtain for the
    Faddeev-Popov determinant. We cannot go further without specifying
    the gauge group. Therefore, for simplicity, in what follows we will
consider
    the group SU(2).  Using the representation (\ref{3.35}) for the matrix
    $V^{ab (Q)}$ and the definition of the operator $\pa ^{-1}$, one  obtains
    \be
    &&\Delta[A^{a (P)}_{-}]=\exp\left(-4DL\int^{g}_{0}dg'
    \pa^{-1}\left( u_2 (|g')D^{(Q)}\right)^{(Q)}(z)|_{z=0}\right),\\
    \label{f22}
    &&\pa^{-1}\left( u_2(|g')D^{(Q)}(z)\right)^{(Q)}|_{z=0} =
    \frac{1}{2g' DL}\left(1-\frac{\sin(g' DL)}{g' DL}\right)
    \eel{f23}
    Thus
    \be
    \Delta[A^{a (P)}_-]=\exp \left\{-2\int_{0}^{gDL}
    x^{-1}\left(1-\frac{\sin x}{x}\right)dx \right\}
    \eel{f24}

    	The Faddeev-Popov determinant depends on the zero modes
    $A^{a (P)}_{-}$ through the upper limit in the integral involved in
(\ref{f24}).
    This means that one does need to introduce the Faddeev-Popov ghosts
    (at least for the group SU(2)), and there is an additional term to the
    self-interaction Hamiltonian of the gluons.

    The integral that appears in the right-hand-side of (\ref{f24}) can
    be  presented in the following form
    \be
    \int_{0}^{z}x^{-1}\left(1-\frac{\sin x}{x}\right)dx=
    -\left(1-\frac{\sin z}{z}\right)+\left(Ci(z)-\ga -\ln z \right)
    \eel{f25}
    where $Ci(z)$ is the cosine integral special function, and $\ga$ is the
    Euler constant. Therefore
    \be
    \Delta[A^{a (P)}_-]&=&
    \exp \left\{ 2\left(1-\frac{\sin(gDL)}{gDL}\right)-2\left(Ci(gDL)-\ga
    -\ln (gDL) \right)\right\}
    \eel{f26}

    \section{Conclusions}

    \hspace{3em}We have considered the light-cone quantization of the
    two-dimensional heavy-quark QCD, explicitly taking into account the zero
    modes contribution
    of the gauge degrees of freedom. We have imposed
    the periodic boundary conditions
    for the gauge degrees of freedom and the antiperiodic ones for the
fermions.
    As ordinary QCD, this model is gauge invariant, meaning that there
    are unphysical degrees of freedom in the problem. We have used the
    Faddeev-Jackiw algorithm to quantize the theory . In order to see
    the role of the zero modes explicitly, we have considered the gauge
conditions
    which are needed only in the  nonzero mode sector. We obtained the physical
    variables (coordinates and their conjugate momenta) and
    the corresponding (anti-) commutation  relations. We found that
    the physical variables
    are the zero modes of the ``spatial" light-cone gauge degrees of freedom
    $A^{a (P)}_{-}$ and $\Pi^{a (P)}_{-}$, and the fermionic variables $\Psi$
    and $ \Pi_{\Psi}$. Solving the constrains and the gauge conditions
    in order to eliminate the unphysical gauge degrees of freedom, we
    constructed the physical Hamiltonian.  In the elimination of
    the unphysical variables mentioned above, we found that the expression
    for the momenta
    $\Pi^{a (Q)}_{-}$  contains an additional term
    that is linear in $x^-$ (see (\ref{3.22})). Such a
    term gives a contribution to  both the self-interaction and
    mutual-interaction potentials of the gluon fields and quark fields.
    We found that
    one needs to introduce the Faddeev-Popov ghosts in their own nonzero
    mode sector. The Faddeev-Popov determinant was calculated, and it was found
    that it depends on the zero modes $A^{a (P)}_{-}$,
    giving a  contribution to the self-interaction Hamiltonian of the gluons.
    \vspace{.3in}

    {\bf ACKNOWLEDGMENT}

    \vskip3ex
    	It is pleasant to thank R.W. Brown, V.K. Mishra and C.C. Taylor
    for useful discussions and remarks that greatly  improved  the
    manuscript. Sh.Sh. is also grateful to E.S. Fradkin for reading the
    manuscript and providing valuable remarks.


\vfill

    \newpage
    \begin{thebibliography}{99}

    \bibitem{brev1}
    S. J. Brodsky, et. al., Part. World {\bf 3}, 109 (1993).

    \bibitem{brev2}
    S. J. Brodsky and H. C. Pauli, ``Light cone quantization of quantum
    chromodynamics", Report No. SLAC-PUB-5558, 1991 (unpublished).

    \bibitem{Brodsky}
     S. J. Brodsky, T. Huang, G. Lepage and P. Mackenzie, in {\it Particles
    and Fields} - 2, Proceedings of the Banff Summer Institute,
    Banff, Alberta, Canada, 1981, edited by A.Z. Capri and A.N. Kamal
    (Plenum, New York, 1983)

    \bibitem{Brodsky1}
     S. J. Brodsky and C. R. Ji, in {\it Quarks and Leptons}, Proceedings of
    the Fourth South African Summer School, Stellenbosch, South Africa,
    1985, edited by C. Engelbrecht, Lecture Notes in Physics, {\bf 248}
    (Springer, Berlin, 1986)

    \bibitem{Pauli}
     H. C. Pauli and S. J. Brodsky, Phys. Rev.{\bf D32}, 1993, (1985); {\bf
    D32}, 2001 (1985).

    \bibitem{Pauli1}
     T. Eller, H. C. Pauli and S. J. Brodsky, Phys. Rev.{\bf D35}, 1493 (1987).

    \bibitem{Pauli2}
     K. Hornbostel, H. C. Pauli and S. J. Brodsky, Phys. Rev.{\bf D41}, 3814
    (1990).

    \bibitem{Pauli3}
     R. J. Perry, A. Harindranath and K. G. Wilson, Phys. Rev.Lett.{\bf 65},
     2959 (1990).

    \bibitem{Pauli4}
     R. J. Perry and A. Harindranath, Phys. Rev.{\bf D43}, 492, 4051 (1991).

    \bibitem{Schlieder}
     S. Schlieder and E. Seiler, Commun. Mathy. Phys. {\bf 25}, 62 (1972).

    \bibitem{Coester}
     F. Coester and W. Polyzou, Report No. ANL-PHY-75240TH-93 (unpublished).

    \bibitem{mccartor}
     G. McCartor, Z. Phys. {\bf C 52}, 611 (1991).

    \bibitem{Heinzl}
     Th. Heinzl, St. Krusche and E. Werner, Phys. Lett.{\bf B256}, 55 (1991);
    Phys. Lett. {\bf B275}, 410 (1992).

    \bibitem{Brown}
    R. W. Brown, J. W. Jun, Sh. M. Shvartsman and C. C. Taylor, Phys. Rev.
    {\bf D48}, 5873 (1993).

    \bibitem{Maskawa}
     T. Maskawa and K. Yamawaki, Progr. Theor. Phys.{\bf 56}, 270 (1976).

    \bibitem{Luscher}M. Luscher, Nucl. Phys. {\bf B219}, 233, (1983).

    \bibitem{Rajeev}S.G.Rajeev, Phys. Lett. {\bf B212}, 203, (1988).

    \bibitem{Hetrick} J. E. Hetrick and Y. Hosotani, Phys. Rev. {\bf D38},
     2621, (1988); Phys. Lett. {\bf D230}, 88, (1989).

    \bibitem{kal1}
    A. C. Kalloniatis and  H. C. Pauli, Z. Phys. {\bf C60}, 255 (1993).

    \bibitem{kal2}
    A. C. Kalloniatis, H.-C. Pauli and S. Pinsky, Phys. Rev. {\bf D50}, 6633,
    (1994).

    \bibitem{Thomas}T. Heinzl. Light front quantization of gauge theories in a
    finite volume. Talk presented at: ``QCD 94", Montpeller, France,
    July 1994 and at International Workshop on ``Light-Cone Quantization and
    Non-Perturbative Dynamics", Poland, August 1994 (unpublished,
hep-th/9409084)

    \bibitem{Dhar}A. Dhar, G. Mandal and S. Wadia, Nucl. Phys. {\bf B436}, 487,
    (1995)

    \bibitem{Tachibana}M. Tachibana. Quantum Mechanics of Dynamical Zero Mode
in
    ${\rm QCD}_{1+1}$ on the Light-Cone. Preprint KOBE-TH-95-01
    (unpublished, hep-th/9504026)


    \bibitem{Dirac}
    P. A. M. Dirac, Canad. J. Math.{\bf 2}, 129 (1950); Lectures on Quantum
    Mechanics (Benjamin, New York, 1964);

    \bibitem{Bergmann}
     P. G. Bergmann, Helv. Phys. Acta, Suppl.{\bf 4}, 79 (1956)

    \bibitem{Casalbuoni}
     R. Casalbuoni, Nuovo Cim. {\bf 33A}, 115, (1976)

    \bibitem{FJ}
    L. D. Faddeev and R. Jackiw, Phys. Rev. Lett. {\bf 60}, 1692, (1988)

    \bibitem{Isgur}
     N. Isgur and M. B. Wise, Phys. Lett.{\bf B232}, 113 (1989).

    \bibitem{joon}
    J. W. Jun and C. Jue, Phys. Rev. {\bf D50}, 2939, (1994).

    \bibitem{Govaerts}
    J. Govaerts, Int. J. Mod. Phys. {\bf A5}, 3625, (1990)

    \bibitem{Berezin}F. A. Berezin, {\it Introduction to algebra and analysis
    with anticommuting variables} (Moscow Univ. Press, Moscow, 1983)

    \end{thebibliography}
    \end{document}